\documentclass[12pt]{article}
\usepackage{epsf}

\newcommand{\bee}{\begin{equation}}
\newcommand{\ee}{\end{equation}}
\newcommand{\bea}{\begin{eqnarray}}
\newcommand{\eea}{\end{eqnarray}}
\newcommand{\R}{\rm I\kern-.2emR}
\newcommand{\C}{\rm \kern.25em\vrule height1.4ex
depth-.12ex width.06em\kern-.31em C}
\newcommand{\N}{{\rm I\kern-.16em N}}
\newcommand{\Z}{{\rm Z\kern-.35em Z}}

\begin{document}
\begin{flushright}
AZPH-TH-97/09 \\
MPI-PhT/97-35 \\
\end{flushright}
\bigskip\bigskip\begin{center}
{\Huge                                                                         
Is the $2D$ $O(3)$ Nonlinear $\sigma$ Model Asymptotically Free?}
\end{center}
\vskip 1.0truecm
\centerline{Adrian Patrascioiu}
\centerline{\it Physics Department, University of Arizona,}
\centerline{\it Tucson, AZ 85721, U.S.A.}
\vskip5mm
\centerline{and}
\centerline{Erhard Seiler}
\centerline{\it Max-Planck-Institut f\"ur Physik}   
\centerline{\it (Werner-Heisenberg-Institut)}
\centerline{\it F\"ohringer Ring 6, 80805 Munich, Germany}
\bigskip \nopagebreak 

\begin{abstract}
We report the results of a Monte Carlo study of the
continuum limit of the two dimensional $O(3)$ non-linear $\sigma$ model.
The notable finding is that it agrees very well with both the
prediction inspired by Zamolodchikovs' $S$-matrix ansatz and with the
continuum limit of the dodecahedron spin model. The latter finding
renders the existence of asymptotic freedom in the $O(3)$ model
rather unlikely.
\end{abstract}
\vskip2mm

Asymptotic freedom is one of the corner stones of modern particle 
physics. This special property  of certain non-Abelian models is supposed 
to explain many remarkable properties of quantum field theory, such as 
the existence of a non-trivial continuum limit, the possibility of grand 
unification of all interactions, etc. This property was discovered in 
1973 in perturbation theory (PT) \cite{gw,pol}. However the correctness of 
this technique within a non-perturbative setting, such as the one offered 
by lattice quantuum field theory, remains mathematically unsettled. In 
the past, we have repeatedly pointed out that there were good reasons to 
doubt the correctness of the PT approach in non-Abelian models 
\cite{pat,f&f,s-i}.

In this letter we would like to report some new numerical results which, 
in our opinion, make the existence of asymptotic freedom in the two
dimensional ($2D$) non-linear $O(3)$ $\sigma$ model extremely unlikely.
Namely we find that the massive continuum limit of this model is, as far 
as we can see, identical to that of the dodecahedron spin model,
which as a discrete spin model must undergo a transition to a phase with 
long range order ({\it lro}) at a finite inverse temperature
$\beta$ and hence it cannot possibly be asymptotically free. The only 
escape out of this bind would be the following scenario: the presently 
observed agreement between the $O(3)$ and the dodecahedron spin models is 
accidental and the phase transition in the dodecahedron model is actually 
weakly first order, a possibility which we cannot rule out, yet for which 
there is no theoretical basis.

The $O(3)$ model is believed to possess only one phase, with exponentially
decaying correlations. It is of
interest because besides being believed to be asymptotically free, it is 
a model for which several theoretical predictions have been made. They 
stem from a conjecture put forward by Zamolodchikov and Zamolodchikov
\cite{zz} regarding its $S$-matrix. Two predictions were made using this
$S$-matrix ansatz:\\
- Hasenfratz, Maggiore and Niedermayer (HMN) \cite{hmn} used it in 
conjunction with the
thermodynamic Bethe ansatz to predict the ratio $\Lambda/m$\\
- Balog, using the ideas of Karowski and Weisz \cite{kw} about 
constructing form factors out of the Zamolodchikov $S$-matrix
\cite{zz}, extended the work of Kirillov and Smirnov \cite{ks} and computed
the current 2-point function \cite{bal} and then later, together with 
Niedermaier and Hauer \cite{bnh}, also the 2-point function of the
energy-momentum tensor as functions of $p/m$. Balog and Niedermaier 
\cite{bn} extended this construction to the two-point functions of the 
spin and the topological density.

While it has been known for several years that the HMN 
prediction for $\Lambda/m$ disagreed with the Monte Carlo value by at
least 15\%, as far as we know no previous work exists comparing the
predictions for the current and spin 2-point functions with Monte Carlo 
data. In this letter we report the results of some Monte Carlo 
investigations of these issues. For $p/m$ less than about 20 we find 
excellent agreement with the theoretical predictions for the current and 
spin 2-point functions. On the other hand, even though we reach 
correlation length $\xi=167$ the value of $\Lambda/m$ disagrees with the
HMN prediction by 15\% and displays an increase with $\xi$, suggesting
that it will not converge to the predicted value.

The numerics were performed using a variant of Wolff's one cluster 
algorithm
\cite{wolff}. One run consisted of at least 100,000 cluster updates used 
for thermalization, followed by at least 7 bins of 100,000 clusters used 
for taking measurements. Thus each run consisted of up to 1,000,000 
clusters used for measurements. This procedure was repeated, starting 
with a different initial configuration; the total number of measurement 
clusters is recorded in Tab.1. The error was estimated in each run from
the bins. If more than one run was performed, we also estimated an
error by treating each run as a bin. Typically the latter error was 
larger and was recorded as the actual error estimate.

The quantities measured were: \\
- Spin 2-point function $G(p)$:
\bee
        G(p)={1\over L^2}\langle |\hat s(p)|^2\rangle;\ \
\hat s(p)=\sum_x e^{ipx} s(x)
\ee
- Current 2-point function $J(p)$:
\bee
        J(p)={1\over L^2}\sum_{a=1}^3\sum_{\mu=1}^2 
\langle |\hat j_\mu^a(p)|^2\rangle;\ \
\hat j_\mu^a(p)=\sum_x e^{ipx} j_\mu^a(x)
\ee
where $j_\mu^a(x)=\beta\epsilon_{abc}s_b(x)s_c(x+\hat\mu)$ \\
- Magnetic susceptibility $\chi$:
\bee
       \chi=G(0)
\ee
- Correlation length $\xi$:
\bee
  \xi={1\over 2\sin(\pi/L)}\sqrt{(G(0)/G(1)-1)}
\ee
We measured these observables in both the $O(3)$ and dodecahedron spin
model. However only in the $O(3)$ model there exists a continuous 
symmetry, leading to a conservation law and to Ward identities \cite{o2}. 
The normalization of the current in eq.(2) was fixed by the latter. In
the dodecahedron spin model there is no a priori conserved current nor 
any Ward identities and it is a remarkable finding of our investigation 
that in the continuum limit the current 2-point function in the 
dodecahedron spin model is identical, up to an overall normalization, 
with that in the $O(3)$ model.

Let us briefly review what is known about the dodecahedron spin model 
with standard nearest neighbour interaction:\\
a) It possesses a high temperature phase with exponential decay of 
correlation functions. \\
b) It possesses a low temperature phase with {\it lro} 
and at least one pure phase for each dodecahedron vertex. \\
c) It seems to have an intermediate phase with algebraic decay
of correlation functions. \\
While the first two properties follow easily from convergent high and low
temperature expansions, the last one has not been proved rigourously. It 
would follow from a rigorous inequality derived by Richard and us 
\cite{prs}, connecting the dodecahedron model with the $Z(10)$ model, 
provided one knew that
\bee
         \beta_m(D)>\beta_c(Z(10)). 
\ee
Here $\beta_m(D)$ is the inverse temperature below which the dodecahedron 
model exhibits {\it lro} and $\beta_c(Z(10))$ the inverse temperature 
above which $Z(10)$ exhibits exponential decay. In ref. \cite{prs2} we 
gave numerical evidence for the existence of such an intermediate 
massless phase. In such a phase, one could expect the discrete icosahedral
symmetry to be enhanced to $O(3)$, just as $Z(N)$ for $N>4$ is enhanced 
to $O(2)$ \cite{FS}. The results reported in this paper regarding the 
similarity of the massive continuum limits of the $O(3)$ and dodecahedron
model suggest that indeed this is the case.

We investigated the massive continuum limit in a thermodynamic volume. 
To be thermodynamic one needs $L/\xi>7$ and all of our runs obyed this 
criterion. To reach the massive continuum limit one takes $\xi\to\infty$ 
keeping $p\xi$ fixed. Since numerically one cannot increase $\xi$ to
$\infty$, one must control lattice artefacts. We did this in two ways:\\
1. For a given action, we studied the approach to the continuum limit at 
fixed $p/m$ as a function of $\xi$ ($m=1/\xi$).\\
2. We studied the dependence of the results upon the lattice action.

For the standard nearest neighbour action we varied the correlation 
length from 11 to 167. Besides this action, we took measurements also
with the `cut-action' ($\beta=0$ but grad($s$) restricted \cite{jsp}) and
with a `Gaussian perfect action' $(s(i)\cdot s(j))$ gets replaced by the 
the square of the angle between the two spins and there is coupling to 
the nearest neighbour and to the site diagonally located \cite{hnperf}). 
Our main conclusion is that the continuum limit is approached more slowly 
at increased $p/m$ and that for instance at $p/m$=50 one needs $\xi$ around
100 for a 1\% deviation in $J(p)$. We observed no striking effect related 
to the lattice action used, only perhaps a slight advantage gained with 
the `Gaussian perfect action' for the dodecahedron spin model (for which 
the Gaussian improvement is totally ad hoc).

To motivate the presentation of our results let us briefly discuss
the form-factor approach of Balog et al. In the spin or current 
2-point function they insert the complete set of $n$-particle
states and write them as an infinite sum, given in terms of form-factors.
$O(N)$ invariance dictates that in $G(p)$ only odd terms contribute, 
while in $J(p)$ only even ones do. Thus $G(p)$ starts with the 1-particle
contribution, which does not depend on the $S$-matrix. Thus, to see the 
dynamics, it is better to subtract this contribution; so we report
\bee
        G_s(x)= [G(x)/G(0)-1/(x^2+1)/1.001687]\times x^2
\ee
Here $x=p/m$, the division by $G(0)$ represents the wave function
renormalization necessary to have a nontrivial continuum limit and the 
numerical factor 1.001687 represents the difference between our and
Balog's normalization of the spin 2-point function \cite{privcom}.
The whole expression was
multiplied by $(p/m)^2$ because, according to renormalized perturbation
theory, after such multiplication, the expression should diverge as 
$\log(p/m)$. The current 2-point function requires no renormalization (in
$O(3)$) and, if one accepts renormalized perturbation theory, is 
expected to diverge as $\log(p/m)$ at large $p$ \cite{bal}. In
the dodecahedron spin model, the normalization of the current 
is arbitrary, hence we could either consider $J(p/m)/J(1)$ and 
compare it with the same quantity measured in $O(3)$ or simply see if 
after multiplication by a constant $J(p)$ in the two models coincide. 

\begin{figure}[htb]
\centerline{\epsfxsize=8.0cm\epsfbox{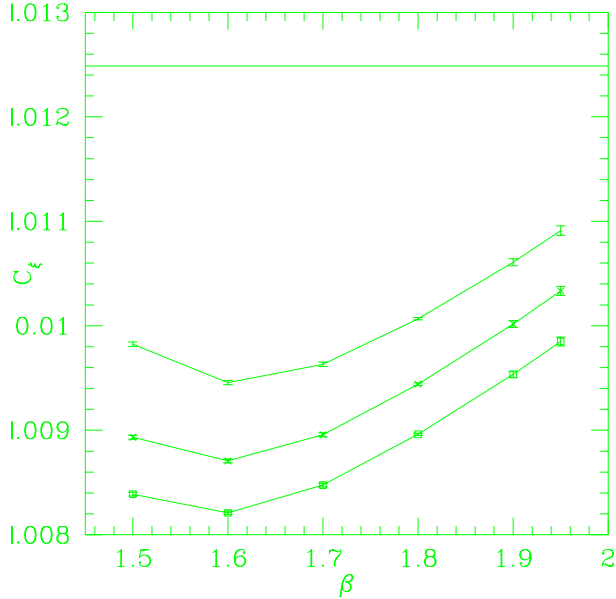}}
\caption{$C_\xi=\Lambda/m$ compared to the HMN prediction (horizontal
line). The data points correspond upwards to 2 loop, 3 loop
and 4 loop PT, respectively.}
\label{xifig}
\end{figure}

\begin{figure}[htb]
\centerline{\epsfxsize=8.0cm\epsfbox{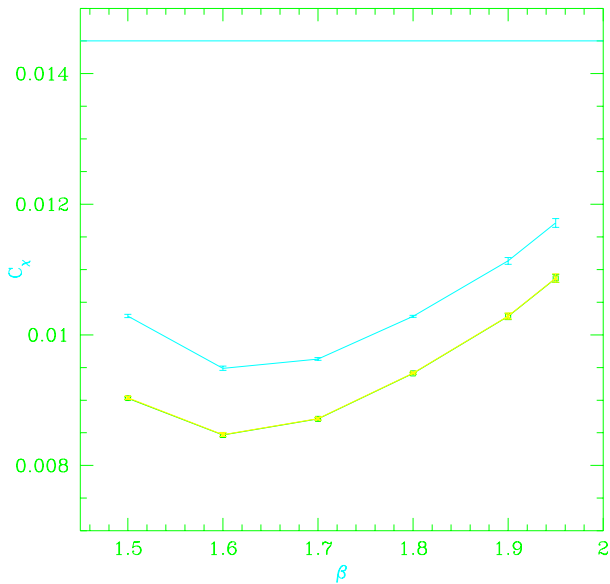}}
\caption{$C_\chi$ compared to the prediction of Alles et al (horizontal
line). The data points correspond upwards to 2 loop and 3 loop
(indistinguishable) and 4 loop PT, respectively.}
\label{chifig}
\end{figure}
{\it Results:}

\begin{figure}[htb]
\centerline{\epsfxsize=8.0cm\epsfbox{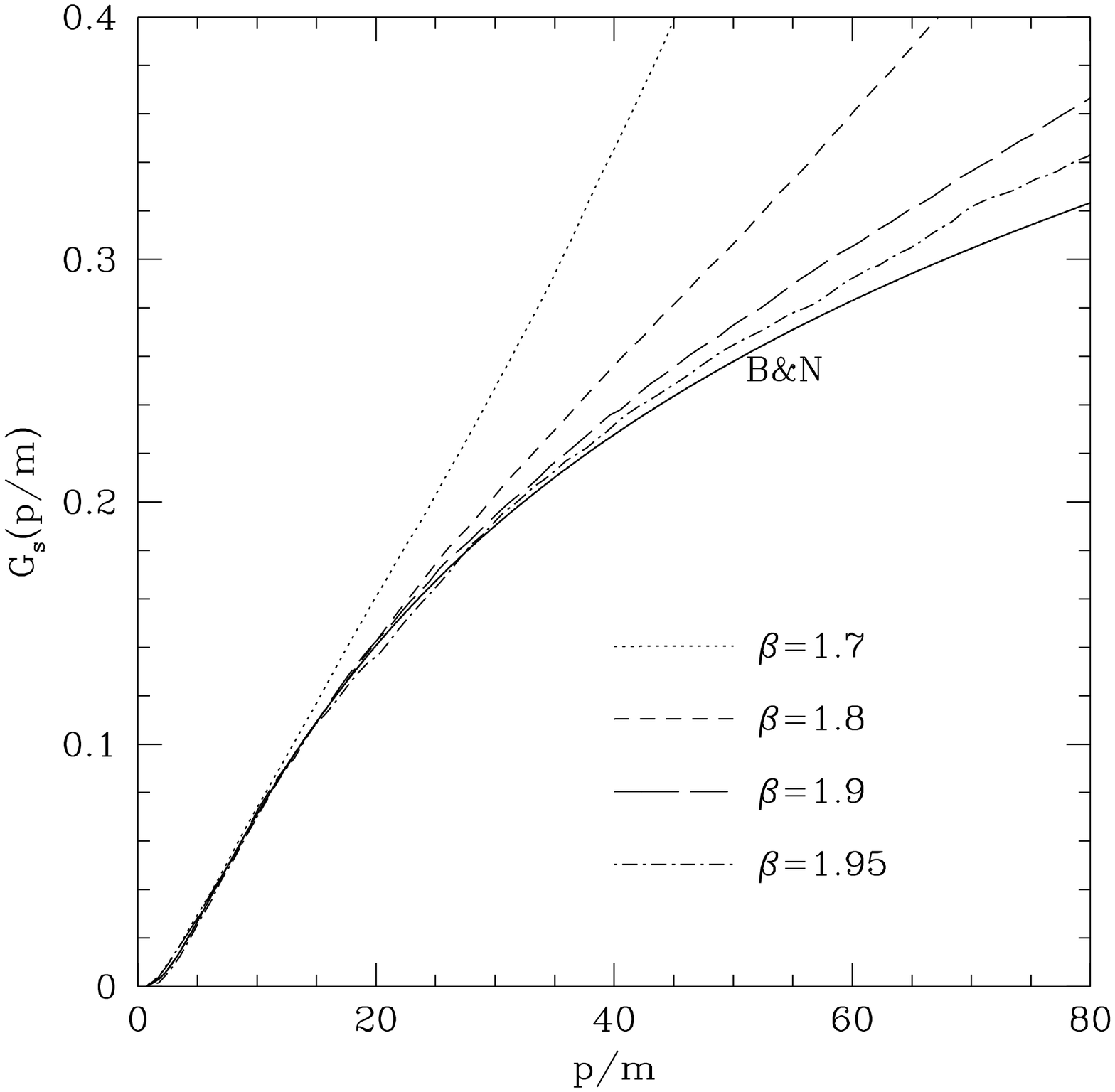}}
\caption{$G_s(p)$ for $O(3)$
as a function of $p/m$. The solid line is the prediction
of Balog and Niedermaier.}
\label{curro3safig}
\end{figure}

\begin{figure}[htb]
\centerline{\epsfxsize=8.0cm\epsfbox{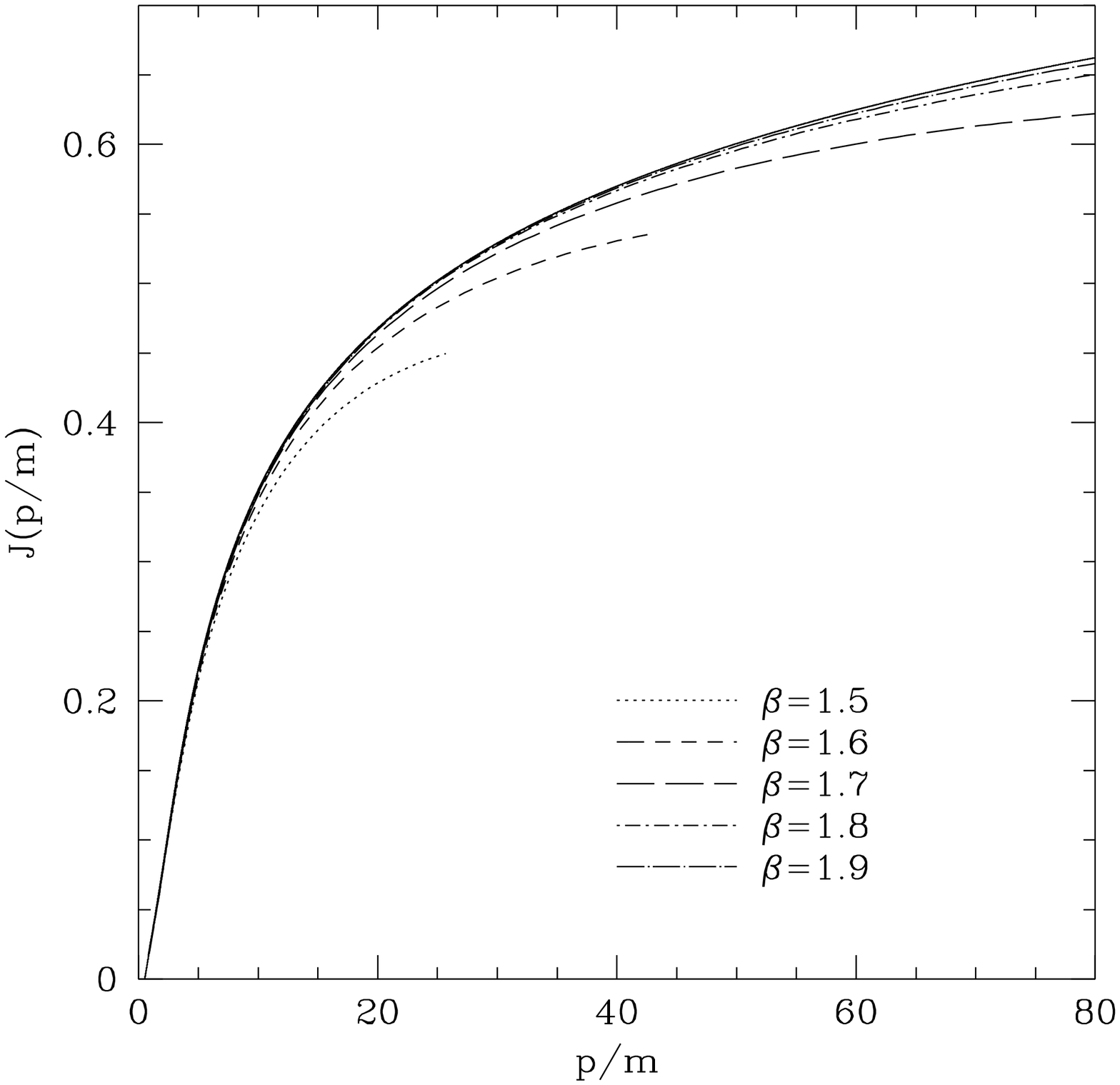}}
\caption{$J(p)$ for $O(3)$
as a function of $p/m$. The solid line is the prediction
of Balog and Niedermaier.}
\label{spino3safig}
\end{figure}

\begin{figure}[htb]
\centerline{\epsfxsize=8.0cm\epsfbox{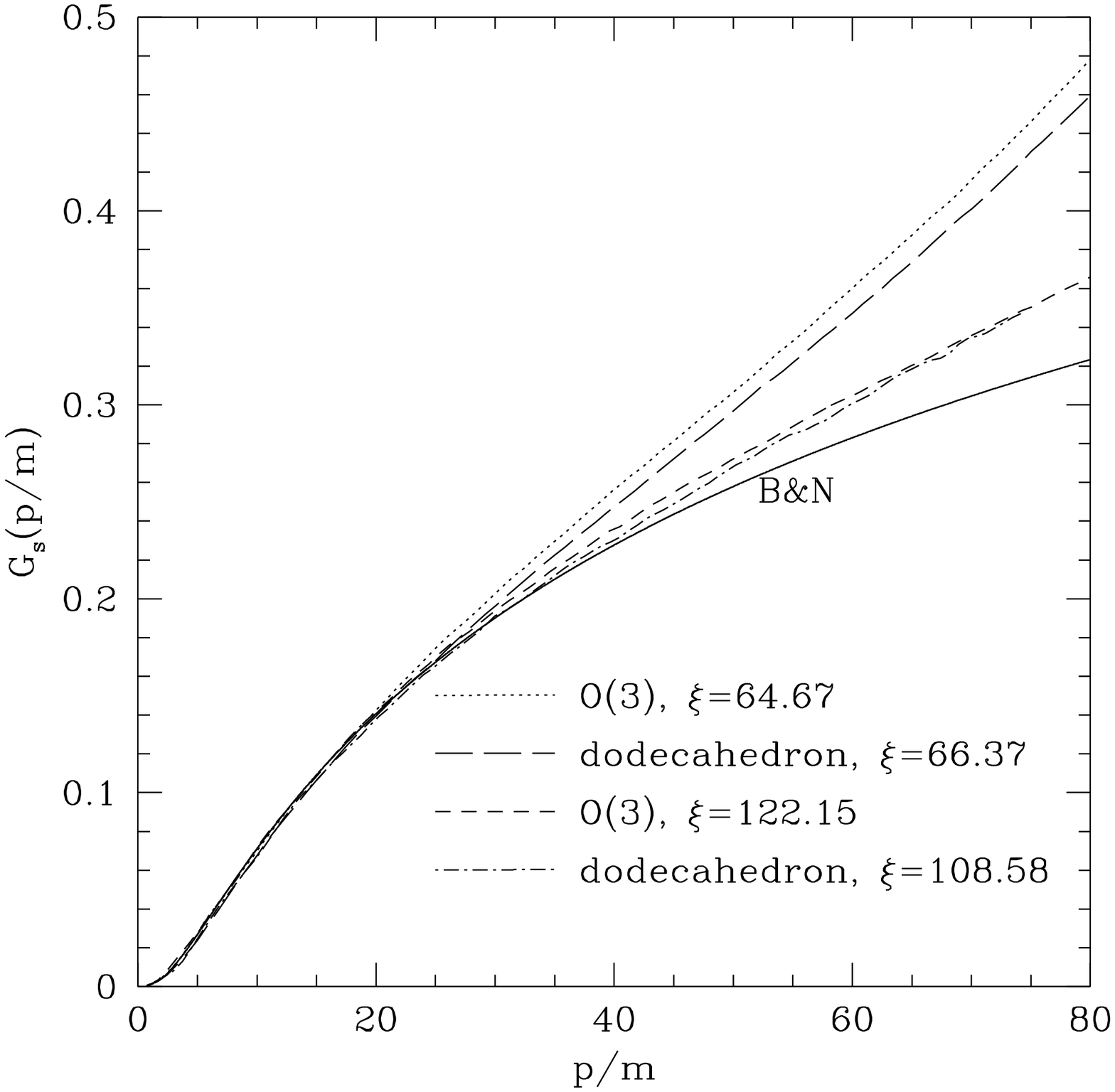}}
\caption{$G_s(p)$ for $O(3)$ and the dodecahedron as a function of $p/m$.
The solid line is the prediction of Balog and Niedermaier.}
\label{spincompfig}
\end{figure}

\begin{figure}[htb]
\centerline{\epsfxsize=8.0cm\epsfbox{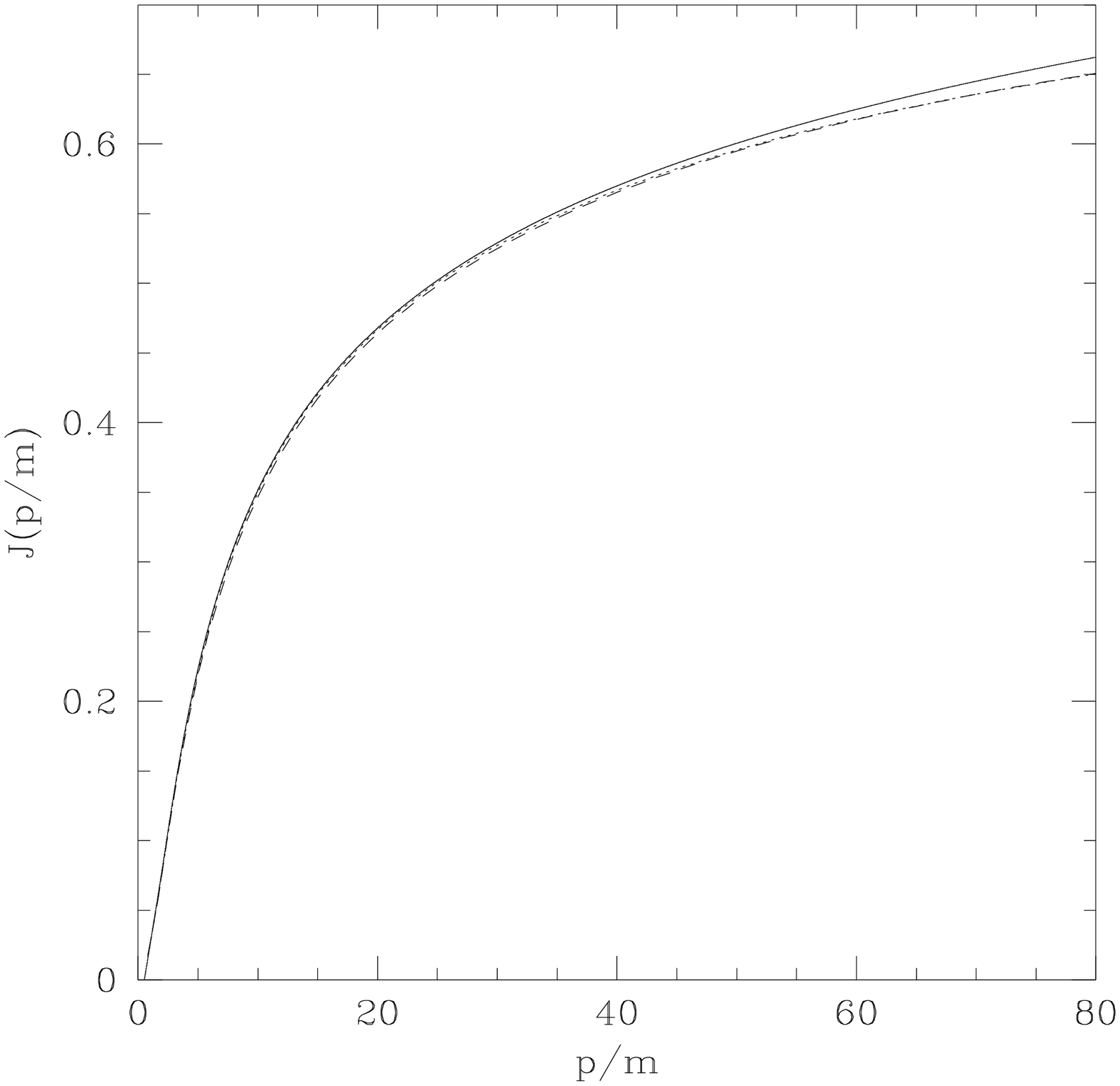}}
\caption{$J(p)$ for $O(3)$ (dotted) and the dodecahedron (dashed)
as a function of $p/m$. The solid line is Balog's prediction.}
\label{currcompfig}
\end{figure}

Perturbative renormalization group arguments lead to the predictions
\bee
\xi\sim {C_\xi e^{2\pi\beta}\over 2\pi\beta}
(1+\sum_{k\geq 1}{a_k\over \beta^k});\ \
\chi\sim {C_\chi e^{4\pi\beta} \over(2\pi\beta)^4}
(1+\sum_{k\geq 1}{b_k\over\beta^k})
\ee
the action dependent coefficients $a_k$ and $b_k$ are known for the 
standard action up to 4 loops \cite{CP}. The HMN prediction is that
\bee
C_\xi={e\over 8} 2^{-5/2} e^{-\pi/2}\approx 0.012487
\ee
Acoording to ref.\cite{alles,CPRV} the HMN prediction implies also
$C_\chi=0.0145$.

In Fig.\ref{xifig}
we present our determination of $\Lambda/m$ in the $O(3)$ model
with standard action, in Fig.\ref{chifig} our determination of $C_\chi$,
compared to the predictions by \cite{hmn} and \cite{alles}, respectively.
The values measured and the number of clusters used 
are recorded in Tab.1.

In Fig.\ref{spino3safig} and \ref{curro3safig}) we compare $G_s(p/m)$ 
(eq.(6)) and $J(p/m)$ (eq.(2)), respectively, for the $O(3)$ model with
standard action at different correlation lengths with the prediction of 
Balog and Niedermaier \cite{bn} and Balog \cite{bal}, respectively.
The errors are less than $10^{-3}$ and therefore not shown.

{\bf Tab.1:} {\it Monte Carlo data for the standard action}

\begin{tabular}[t]{r|r|r|r|r}
$\beta$ & $L$ & clusters & $\xi$& $\chi$ \\
\hline
\hline
1.5 & $ 78 $ & $1\times 10^6$ & 11.03(2) & 175.75(43) \\
\hline
1.6 & $140 $ & $1\times 10^6$ & 18.97(4) & 447.30(6) \\
\hline
1.7 & $250 $ & $2\times 10^6$ & 34.55(7) & 1268.(3.) \\
\hline
1.8 & $500 $ & $4\times 10^6$ & 64.67(8) & 3831.(6.) \\
\hline
1.9 & $910 $ & $2\times 10^6$ & 122.15(39)& 11847.(67.)\\
\hline
1.95& $1200$ & $7\times 10^5$ & 168.39(71)& 21148.(123.)
\end{tabular}\\

In Fig.\ref{spincompfig} (\ref{currcompfig}) we compare $G_s(p/m)$ ($
J(p/m)$) for the $O(3)$ and the dodecahedron models, both with standard action.
For the dodecahedron model $J(p/m)$ was multiplied by a renormalization factor
determined by fitting.

{\it Discussion:}

1. The data in Fig.1 agree very well with prior data \cite{wo,apos} and 
indicate that most likely $\Lambda/m$ will increase past the 
HMN prediction. Attempts to improve this situation by 
using the so-called `energy scheme` \cite{alles} produced similar results
(the curve is shifted closer to the predicted value but it shows a 
non-vanishing slope away from it).

2. A comparison of $\chi(\beta)$ with the predictions of asymptotic
freedom shows the same behavior: the data increase faster than expected, 
in agreement with the existence of a critical point at finite $\beta$.
From the data in Tab.1 we find that $C_\chi$ is growing steadily, so far
reaching 0.0111, compared to the `exact' prediction of a constant
value of 0.0145.


3. The numerical agreemeent of $G_s(p/m)$ and (the normalization
adjusted) $J(p/m)$ in the $O(3)$ and dodecahedron spin model and their 
excellent agreement with the Balog and Balog-Niedermaier prediction 
strongly suggest that the two models possess the same continuum limit, 
which is correctly predicted by the Zamolodchikovs $S$-matrix ansatz. 
Since
the massive phase of the dodecahedron model terminates at finite $\beta$, 
its continuum limit cannot have asymptotic freedom. Indeed, for instance
the running coupling $g_r(L)$ introduced in \cite{lww} cannot vanish at
short distances, because at any $\beta$: (a) $g_r(L)_{O(2)}$ does not vanish 
and (b) $g_r(L)_{O(3)}\geq g_r(L)_{O(2)}$ \cite{kpv}.

4. Balog and Niedermaier\cite{bnscal} have put forth an ansatz
for the contribution of higher $n$-particles to $G(p/m)$ and
$J(p/m)$. According to them this ansatz would explain why the 
contribution of $n>6$ is negligible at $p/m<10^4$ and how the same
contributions sum up at asymptotic $p/m$ to produce the
logarithmic increase of both $G_s(p/m)$ and $J(p/m)$ corresponding to
asymptotic freedom. Our data suggest that their ansatz is inadequate, since 
the agreement of $O(3)$ with the dodecahedron model rules out such a 
logarithmic increase in $J(p)$ (see \cite{o2}).

AP is grateful to the Humboldt Foundation for a Senior US Scientist
Award. We acknowledge numerous discussions with J.Balog and 
M.Nie\-der\-mai\-er.

\end{document}